\documentstyle{mn2e}
\input psfig.sty 

\def\solmas{{M$_\odot$}}
\def\simless{\mathbin{\lower 3pt\hbox
   {$\rlap{\raise 5pt\hbox{$\char'074$}}\mathchar"7218$}}}   
\def\simgreat{\mathbin{\lower 3pt\hbox
   {$\rlap{\raise 5pt\hbox{$\char'076$}}\mathchar"7218$}}}   
\def\etal{{\rm et al.}}

\def\solmas{{M$_\odot$}}
\def\solm{{M_\odot}}

\def\au{ AU}

\def\tff {t_{\rm ff}}

\def\AaA{{ A\&A}}

\def\ApJ{{ApJ}}
\def\apj{{ApJ}}
\def\apjl{{ApJL}}

\def\mnras{{MNRAS}}

\def\ARAA{{ARA\&A}}

  \makeatletter
  \ifx\CUP@mtlplain@loaded\undefined  
  \makeatother  
    
  \else
    
  \fi

\loadboldmathitalic 
\loadboldgreek      
  
\makeatletter
\ifx\CUP@mtlplain@loaded\undefined  
  \newfont\bit{cmbxti10 at 9pt}
\else
  \newfont\bit{mtbxti10 at 9pt}
\fi
\makeatother

\def\LaTeX{L\kern-.36em\raise.3ex\hbox{a}\kern-.15em
    T\kern-.1667em\lower.7ex\hbox{E}\kern-.125emX}

\newcommand{\gsim}{\mathrel{\hbox{\rlap{\lower.55ex \hbox {$\sim$}}
                   \kern-.3em \raise.4ex \hbox{$>$}}}}
\newcommand{\lsim}{\mathrel{\hbox{\rlap{\lower.55ex \hbox {$\sim$}}
                   \kern-.3em \raise.4ex \hbox{$<$}}}}

\title[Massive star formation] {Massive star formation: Nurture, not nature}
\author[I. A. Bonnell \etal]
  {Ian A. Bonnell$^1$\thanks{E-mail: iab1@st-and.ac.uk}, Stephen G. Vine$^1$ \& Matthew R. Bate$^2$\\ 
$^1$ School of Physics and
  Astronomy, University of St Andrews, North Haugh, St Andrews, Fife,
  KY16 9SS. \\
$^2$ School of Physics, University of Exeter, Stocker
  Road, Exeter EX4 4QL \\ }

\date{\today}

\begin{document}

\maketitle

\begin{abstract}
We investigate the physical processes which lead to the formation of
massive stars. Using a numerical simulation of the formation of a
stellar cluster from a turbulent molecular cloud, we evaluate the
relevant contributions of fragmentation and competitive accretion in
determining the masses of the more massive stars. We find no
correlation between the final mass of a massive star, and the mass of
the clump from which it forms.  Instead, we find that the bulk of the
mass of massive stars comes from subsequent competitive accretion in a
clustered environment. In fact, the majority of this mass infalls onto
a pre-existing stellar cluster. Furthermore, the mass of the most
massive star in a system increases as the system grows in numbers of
stars and in total mass.  This arises as the infalling gas is
accompanied by newly formed stars, resulting in a larger cluster
around a more massive star.  High-mass stars gain mass as they gain
companions, implying a direct causal relationship between the cluster
formation process, and the formation of higher-mass stars therein.
\end{abstract}


\section{Introduction}

There are currently two competing ideas as to how massive stars form.
Do they form as essentially a scaled-up version of low-mass stars
(e.g. Shu, Adams \& Lizano~1987), where the star's mass is determined
by the mass in the molecular core that collapses due to its
self-gravity? Alternatively, is the final mass of massive stars
determined through environmental processes such as competitive
accretion or mergers in clusters? In the first scenario, the stellar
environment is unimportant and the mass is decided by the amount of
mass which is necessary to be self-gravitating considering either
thermal or turbulent support (Yorke \& Kruegel~1977; McKee \&
Tan~2002,~2003).  Thus, high-mass clumps in molecular clouds, if they
are just self-gravitating, lead directly to high-mass stars (Padoan \&
Nordlund~2002).  This is an attractive possibility as the high-mass
tail of the clump-mass spectrum resembles the stellar initial mass
function (eg., $\rho$ Ophiuchus;  Motte, Andr\'e \& Neri~1998). McKee
\& Tan~(2003) envision that the pre-massive star clumps are located in
the high pressure regions in the centre of clusters, although such a
spatial segregation of massive clumps is not found in the $\rho$
Ophiuchus and Serpens protoclusters (Elmegreen \& Krakowski~2001).

In the second scenario, the star's mass is strongly influenced, if not
determined, by the stellar environment, a stellar cluster. Individual
stars compete for the reservoir of gas (Zinnecker~1982; Larson~1992),
with those sitting in the bottom of the potential winning the
competition and thus accreting more gas and attaining higher masses
(Bonnell \etal~1997;~2001a). In this case, the final masses need have
little correlation with their initial masses, the accretion can
explain the full range and distribution of stellar masses (Bonnell
\etal~2001b). A similar process was discussed by Murray \& Lin~(1996)
where gas parcels were evolved under purely dynamical forces
until they collided and merged. Parcels that become gravitationally unstable 
collapse to form stars and grow in mass through further collisions with
other gas parcels.

In both cases, the massive star grows through accretion onto
a lower-mass protostellar core
(Behrend \& Maeder~2001). One potential difficulty with accretion
 is that feedback in the form of radiation
pressure may limit the mass accumulation process to masses of order
$\sim 10$ \solmas (Wolfire \& Cassinelli~1987; Yorke \& Kruegel~1977; Edgar \& Clarke~2003), although a rapid rotation
may sufficently reduce the star's luminosity in the equatorial plane (Yorke
\& Sonnhalter~2002). Thus,  more exotic processes such as stellar mergers
may be potentially required in order to explain the formation of massive
stars (Bonnell, Bate \& Zinnecker~1998).

Observationally, there are a number of circumstantial clues that
support an environmental influence on massive star
formation. Massive stars are even more likely than their low-mass
counterparts to be found in stellar clusters (Clarke, Bonnell \&
Hillenbrand~2000; Lada \& Lada~2003). Young clusters are generally
mass segregated with the most massive stars found in their
cores. Additionally, there is an observational correlation of the mass
of intermediate and high-mass stars with the stellar density of the
surrounding cluster (Testi, Palla \& Natta~1999; Hillenbrand~1995, see
Clarke \etal~2000). For example, Testi \etal~(1997,~1999) surveyed the
environments of pre-main sequence Herbig AeBe stars and found that
stars more massive than $\approx 6 \solm$ are generally surrounded by
a cluster of lower-mass stars and that the number of stars in the
cluster increases with increasing mass.  This implies a potential
causal relationship between the number of stars in the cluster and the
mass of the most massive star. The Herbig AeBe data is also consistent
with random pairing from an initial mass function (IMF) (Bonnell \&
Clarke~1999) although this would not explain the observed mass
segregation in larger clusters.

In this paper, we investigate the formation of massive stars that
occur in a numerical simulation of the fragmentation of a turbulent
molecular cloud and the subsequent formation of a stellar cluster
(Bonnell, Bate \& Vine~2003). This simulation showed that the stellar
cluster, containing approximately 400 stars, formed through a
hierarchical fragmentation and merging process. The turbulence leads
to several sites of star formation. Individual stars form in
filamentary structures and then fall towards their local potential
minima. This forms small-N subclusters which grow through accreting
infalling stars and gas. The subclusters eventually merge, aided by
the dissipation of kinetic energies by the gas, to form one large
cluster. The simulation formed six stars with masses greater than 10
\solmas\ with a maximum stellar mass of $\approx 27$ \solmas, ignoring any radiative feedback (e.g. Yorke \& Kruegel~1977; Yorke \& Sonnhalter~2002). In
section 2 we detail the calculations.  Section 3 investigates the
origin, in the simulation, of the massive stars.  Section 4 relates
the massive star formation to the process of cluster formation and the
resultant cluster properties. We discuss the implications of this work
for massive star formation in Section 5 while our conclusions are
given in Section 6.

\section{Calculations}

The results presented in this paper are based on a numerical
simulation performed with the Smoothed Particle Hydrodynamics (SPH)
method (Monaghan~1992, Benz \etal~1991). The details of this
simulation have already been presented in Bonnell \etal~(2003), and we
summarise the relevant details here. The simulation followed the
fragmentation of a turbulent molecular cloud containing 1000 \solmas\
in a region of 0.5 pc radius for $2.5 t_{ff}$ or $4.75 \times 10^5$
years. The gas is isothermal at $10$ K as expected for densities
$\simless 10^{-13}$ g cm$^{-3}$ (eg., Larson~1969; Masunaga, Miyama \&
Inutsuka~1998). The supersonic turbulence is modelled by including a
divergence-free random Gaussian velocity field with a power spectrum
$P(k) \propto k^{-4}$ where $k$ is the wavenumber of the velocity
perturbations (Ostriker, Stone \& Gammie~2001).
The dependence of the fragmentation and resultant IMF
has been shown to be rather insensitive to the slope of the
power spectrum (Delgado-Donate, Clarke \& Bate~2004; cf Klessen, Heitsch and Mac Low~2000 and Klessen \&
Burkert~2001). In three dimensions,
this matches the observed variation with size of the velocity
dispersion found in molecular clouds (Larson~1981). The velocities are normalised to make the kinetic
energy equal to the absolute magnitude of the potential energy so that
the cloud is marginally bound.  In contrast, the thermal energy is
initially only 1 per cent of the kinetic energy. The Jeans mass of the
cloud is then $1 \solm$, and the cloud contains 1000 thermal Jeans
masses. The turbulence leads to the generation of shocks, structure and the dissipation of kinetic energy (Mac Low \etal~1998, Ostriker \etal~2001).
We do not include any feedback (radiative or kinematic) from
the newly formed stars, which is a limitation as feedback from massive
stars can be dynamical important (radiation pressure, HII regions etc).
Osorio, Lizano \& D'Alessio~(1999) discuss how the incipient HII
region from a young massive star can be choked off through high
accretion rates.
The simulation was carried out on the United Kingdom's Astrophysical
Fluids Facility (UKAFF), a 128 CPU SGI Origin 3800 supercomputer.

Dense protostellar fragments are replaced by sink-particles in order
to follow the evolution further (Bate, Bonnell \& Price~1995).  These
sink-particles accrete infalling gas that falls within a sink-radius
of 200 \au\ if they are bound to the sink-particle, whereas all gas
particles that fall within 40 \au\ are accreted, regardless of their
properties.  The simulation used $5\times 10^5$ particles, implying a
minimum protostellar mass of $0.1$ \solmas (e.g. Bate \&
Burkert~1997). Fragments with lower-masses are not resolvable in this
simulation.  The gravitational accelerations between sink-particles
are smoothed within distances of 160 \au.  Stellar collisions are not included in the simulation.

\section{The origin of massive stars}

\begin{figure}
\centerline{\psfig{figure=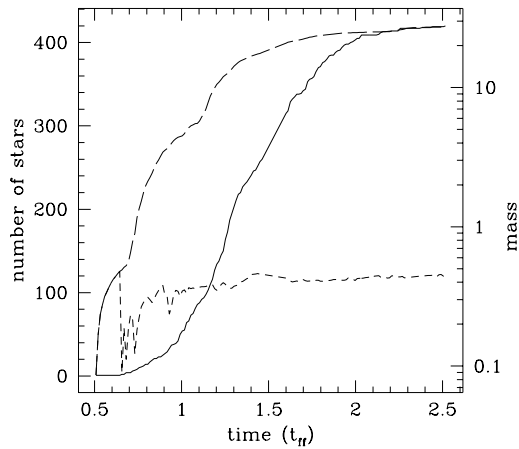,width=3.0truein,height=3.0truein}}
\caption{\label{NMevol} The evolution of the number of stars (solid line),
the maximum stellar mass (long-dashed line) and median stellar mass (short-dashed line) are plotted as a function of time in units of the free-fall time
($t_{ff} = 1.9 \times 10^5$ years).
The masses are in \solmas\ and should be read against the axis on the right
side. Note that the same star is not always the most massive star present.}
\end{figure}

\begin{figure*}
\centerline{\psfig{figure=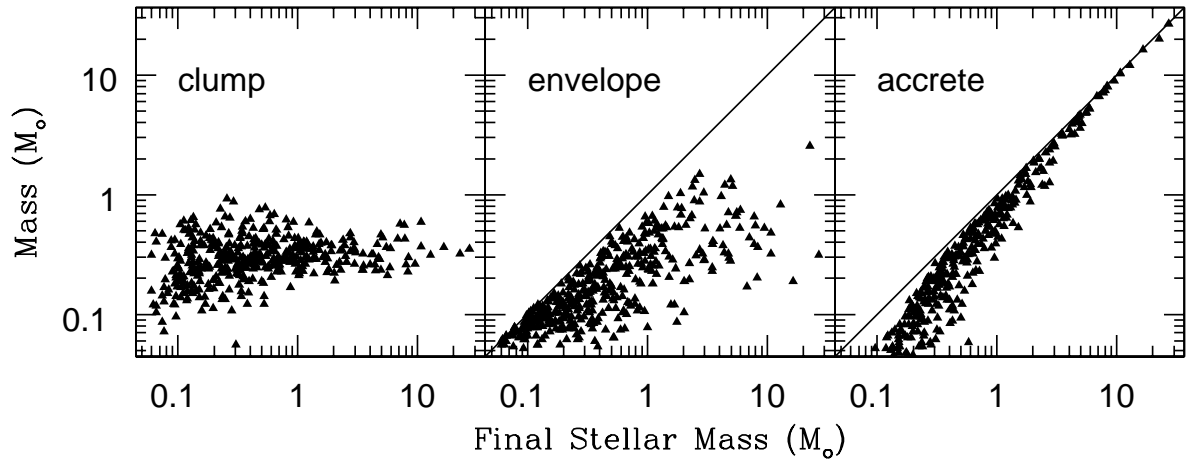,width=7.5truein,height=7.5truein,rwidth=7.5truein,rheight=3.05truein}}
\caption{\label{initvsfinmass} Three potential sources for the stellar masses,
the clump structures in the cloud (left panel), the surrounding envelopes 
(middle panel) and subsequent accretion (right panel), 
are plotted against the final stellar masses (all in \solmas).
The clump masses and envelope masses (see text) are determined at the time of protostar formation. We see no correlation
between the clump masses and the final masses. Instead, the
masses of lower-mass ($m \simless \solm$) stars are largely due to the envelope masses
at the time of protostar formation. In contrast, the formation of
higher-mass stars (with $m \simgreat  \solm$) requires subsequent accretion 
in order to explain their masses. The solid lines in the middle and right
panel indicate a one to one relation with the final mass.}
\end{figure*}

One of the major results of the simulation reported in Bonnell
\etal~(2003) is that the cluster formation process naturally results
in a distribution of stellar masses from $0.1$ to $\approx 27$ \solmas, with a median mass of $\approx 0.43$
\solmas, and that agrees broadly with observed initial mass functions.
Figure~\ref{NMevol} shows the evolution of the number of stars present in the
cluster as well as the maximum and median stellar masses. The median
mass is approximately constant throughout the evolution while both the number
of stars, and the maximum  stellar mass, increase. 

This result could be misinterpreted as being due to the random
sampling from an IMF such that when more stars are present, the chance
of having a more massive star increases. This neglects the fact that
the individual stars accrete mass throughout the evolution. Thus for
example, the  star that is the most massive star (at 5 \solmas) 
when $\approx 50$ stars are present, is the same star that ultimately 
is the most massive star (at
$\approx 30 \solm$) in the final cluster of $\approx 400$ stars. 

It is
not always the same star which is the most massive star in the system.
The discontinuities in the maximum stellar mass indicates where
different stars take over being the most massive star present.  We
thus have a self-consistent star formation laboratory with which to
study massive star formation, and how individual stellar masses are
determined. For example, we can investigate the clump-masses from
which the stars form and thus whether the physical conditions of the
pre-collapse clumps determines the stellar masses. Furthermore, SPH
being a Lagrangian method, we can trace back where the mass, that
ultimately comprises the more massive stars, originates in the
molecular cloud.

We estimate the clump-mass from which each star forms, at the time of
formation, as the mass contained within a spherical radius where the
local gas density is continually decreasing. The end of a clump is
then defined as when the local gas density starts to rise, or when
another star is encountered. We see
in figure~\ref{initvsfinmass}
that there is no correlation between these clump masses and the final
stellar masses. The clump masses extend from above our resolution
limit of $0.1 \solm$ \ to $\approx 1 \solm$. The most massive stars
originate from clumps that are near the median of the distribution at
$\approx 0.4 \solm$, which is itself approximately the final median stellar mass of
$\approx 0.43$ \solmas. Some of the more massive clumps actually result
in lower-mass stars suggesting either multiple fragmentation or that a
significant fraction of the clump is accreted by another star.

\begin{figure*}
\centerline{\psfig{figure=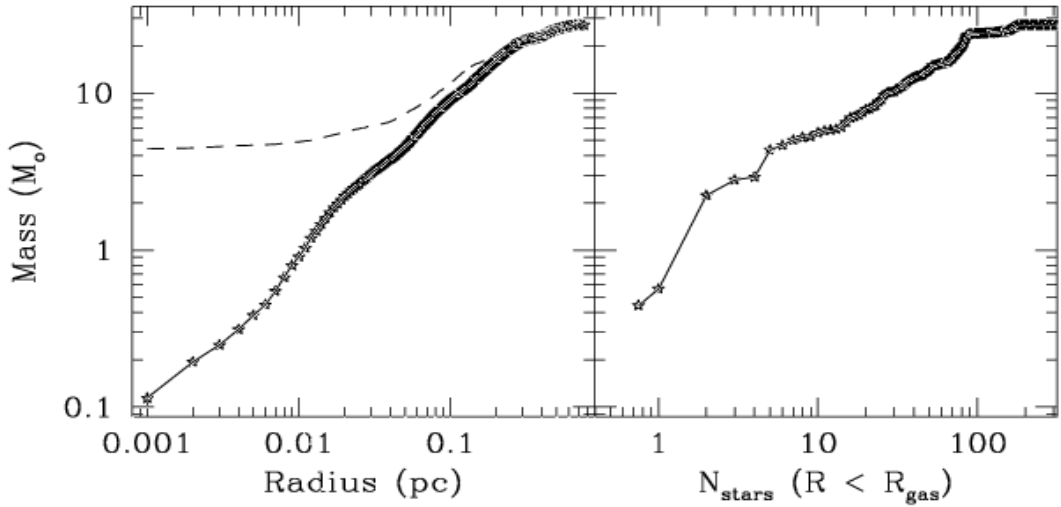,angle=270.0,width=6.0truein,height=3.05truein,rwidth=6.0truein,rheight=3.05truein}}
\caption{\label{massaccprof} The  cumulative mass distribution of the gas
that ultimately accretes onto the most massive star  is plotted as a function of its distance from the star at the time of formation (solid line) in the left hand panel. The distribution when this star has 10 lower-mass companions within $0.1$ parsecs is also plotted (dashed line). The same distribution is plotted in the right-hand panel against the maximum number of other stars contained within this gas (closer to the target star).
Thus, little gas is able to accrete before additional stars are present
to compete for the mass reservoir.}
\end{figure*}

\begin{figure*}
\centerline{\psfig{figure=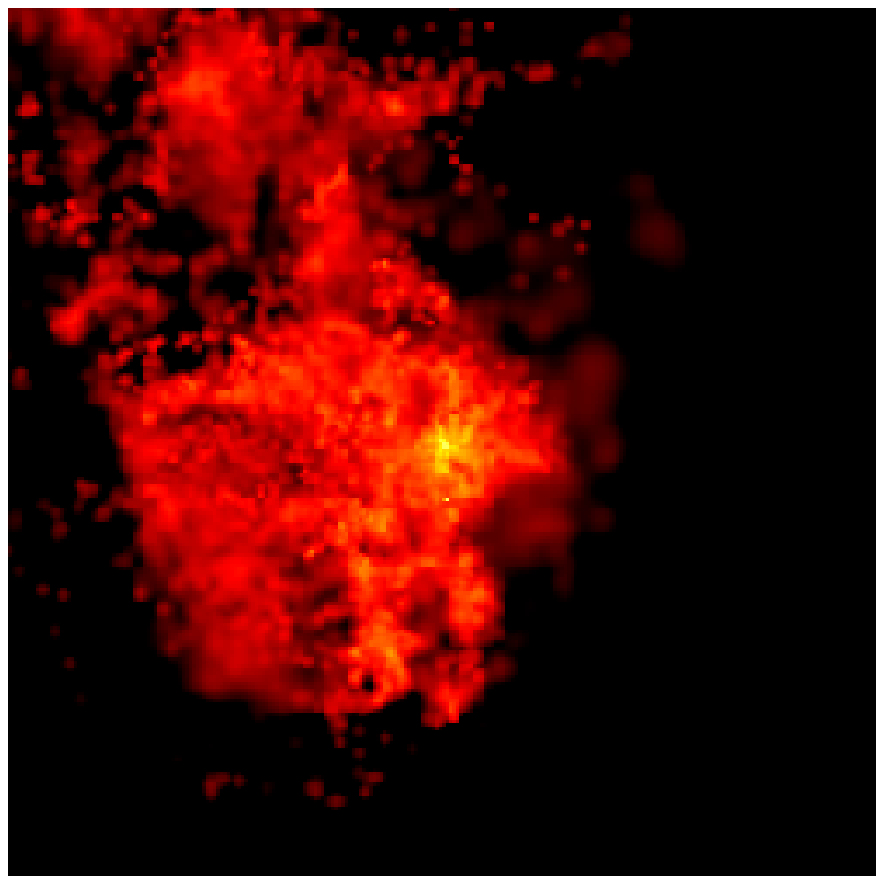,width=3.20truein,height=3.20truein}
\psfig{figure=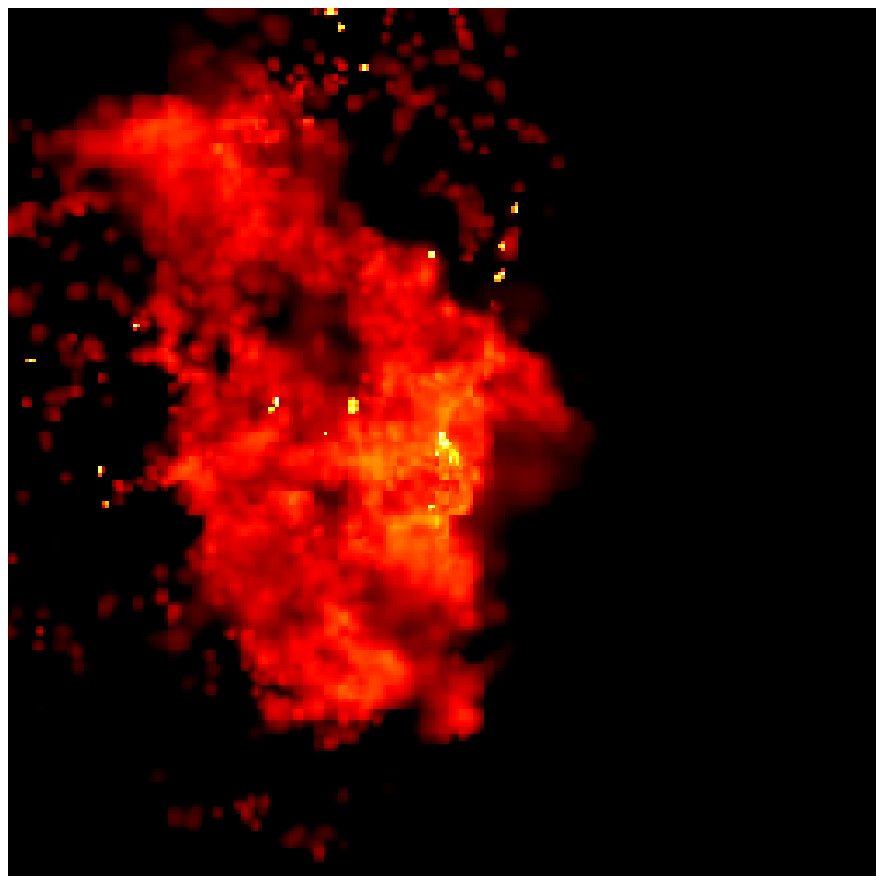,width=3.20truein,height=3.20truein}}
\centerline{\psfig{figure=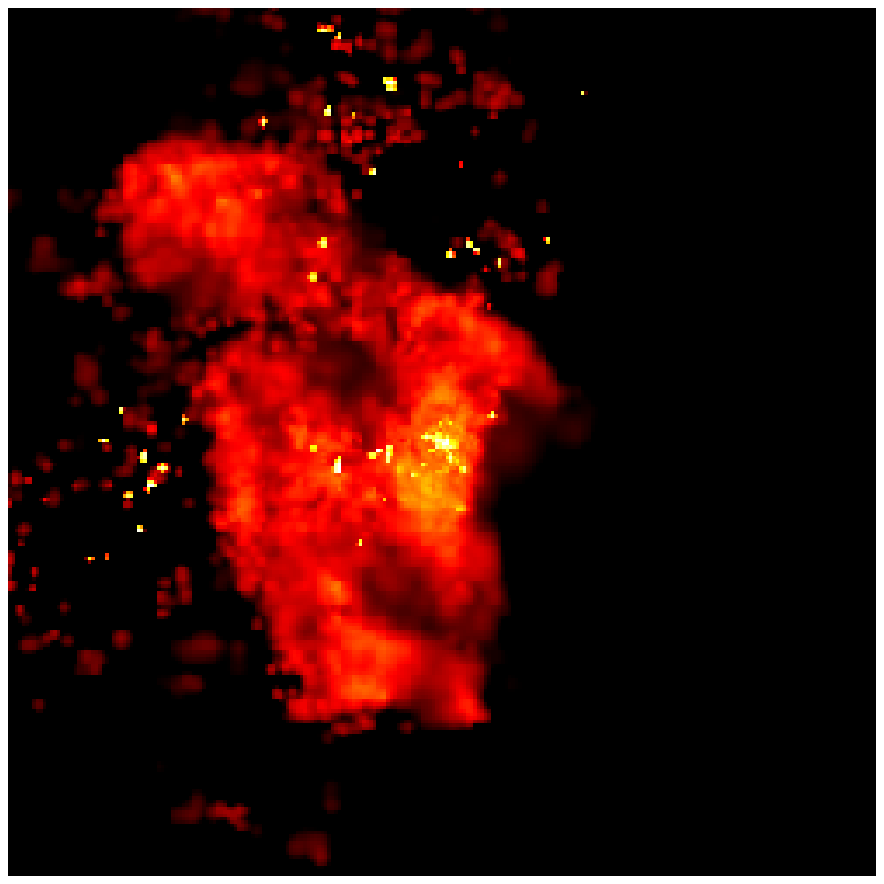,width=3.20truein,height=3.200truein}
\psfig{figure=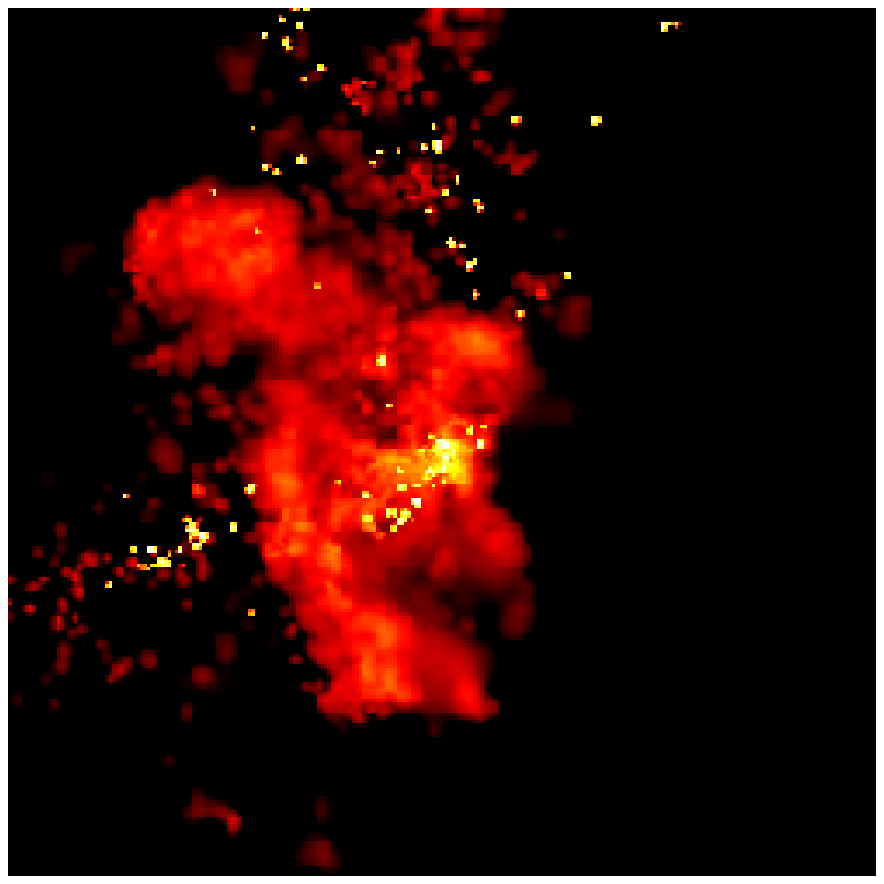,width=3.20truein,height=3.200truein}}
\caption{\label{acchist} The spatial distribution of the column
density of the gas that accretes onto the most massive star is shown
at four different times.  Each panel shows a region $0.8$ parsecs on a
side centred on the star concerned. The upper-left panel shows the
distribution at the time the star forms ($t=0.66 t_{ff} = 1.26 \times
10^5$ years).  Note there is only one pre-existing star. The
upper-right panel shows the distribution when the star has 10
companions within $0.1$ parsecs and a mass of $4.42 \solm$ ($t=1
t_{ff} = 1.9 \times 10^5$ years) while the lower-left panel shows the
distribution when the star has 25 companions within $0.1$ parsecs and
a mass of $5.86 \solm$ ($t=1.15 t_{ff} = 2.19 \times 10^5$ years). The
lower-right panel shows the distribution once the star has accreted up
to $8 \solm$ and is now part of a rich cluster containing many tens of
stars ($t=1.3 t_{ff} = 2.46 \times 10^5$ years).}
\end{figure*}

As mentioned above, we can use the Lagrangian nature of SPH to
reconstruct the mass accretion history for each star and determine
where the mass originated.  The middle panel of
figure~\ref{initvsfinmass} plots the relation between the final
stellar masses and the mass contained within an initial envelope
around the forming star. This envelope is defined as a spherical
region in which at least 99 per cent of the gas ultimately ends up on
the star concerned. This envelope represents
a mass reservoir from which the star accretes unimpeded by the
presence of other stars. Thus, although it includes mass added through
accretion after protostar formation, it excludes any mass added through
competitive accretion in a cluster environment.   
This envelope mass can account for some of the lower-mass stars but 
fails to recover the masses of the more massive stars ($\simgreat 2 \solm$). 
Even though one
of the more massive stars has the most massive envelope, this envelope
represents only $\approx 10$ per cent of the final mass.  
There is a definite trend that higher-mass stars originated within more massive envelopes. An initially higher-mass aids in the subsequent accretion process.
The
right-hand panel of figure~\ref{initvsfinmass} plots the remaining
mass that therefore is accreted from regions outside this envelope and
therefore from regions which contribute mass to more than one star.
This accretion accounts for the 
bulk of the mass of more massive stars.
Thus we see the role played by competitive accretion far outweighs that
of the structure in the molecular cloud in determining the masses of higher-mass stars. 

We gain a better understanding of where the mass of the more massive
stars comes from in figure~\ref{massaccprof} which plots the
cumulative distribution of the mass, which ultimately comprises the
most massive star, at the time of protostar formation.  The mass
distribution extends throughout the cloud, with only about a third of
the total mass initially contained within $0.1$ parsecs.
Figure~\ref{massaccprof} also plots the cumulative mass distribution
at a time when it has 10 other stars as companions within $0.1$
parsecs, the typical size of the sublcusters that form in the
simulation. Even once the star has 10 other stars in its subcluster,
the bulk of the star's mass is at large distances and has to be
accreted from outside the system. This can be seen from
Figure~\ref{acchist} that shows the spatial distribution of the mass
that ultimately is accreted by this star at four different epochs,
protostar formation ($t=0.66 t_{ff}$, $t_{ff}=1.9 \times 10^5$ years), when the star has 10 and 25 neighbours within
$0.1$ parsecs (at times $1$ and $1.15 t_{ff}$ and masses of $4.42$ and
$5.86 \solm$), and when the star attains a mass of $\approx 8 \solm$
(at $t=1.3 t_{ff}$).  

At the time of protostar formation (upper-left
panel), there is only one other star present in the system and there
is a well defined clump around the forming star. The gas is extended
over much of the volume from which the cluster as a whole ultimately
forms. The mass distribution is still very extended when 10 and 25
companion stars are contained within $0.1$ parsecs. By the time the
star has accreted up to $\approx 8 \solm$, the gas distribution which
makes up the remaining 19 \solmas\ is still very extended, even though
many other stars have since formed which make up the cluster as a
whole (Bonnell \etal~2003).  

The right-panel of
figure~\ref{massaccprof} plots the distribution of the accreted mass
against the maximum number of stars that lie within this gas and its
target star. We can see from this figure that the vast majority of the
gas comes from outside a significant group of stars and could in
principle be accreted by any of them. That this one star is able to
accrete such a significant fraction of the infalling gas is due to the
nature of competitive accretion in clusters. The accretion rate is
then determined by a combination of the star's mass and kinematics
such that more massive stars, which generally move slower, accrete at
significantly higher rates than do lower-mass stars (Bonnell
\etal~2001a).

In general, the stars that eventually attain higher masses
form earlier in the simulation. This provides more time for them to accrete, but more importantly, they have less competition initially such that they are already more massive than the average star by the time they have many companions.
They are then more likely to accrete enough gas to become massive stars. Thus,
five of the six highest mass stars (with $m \simgreat 10 \solm$) form out of the initial 50 stars, but there remain many low-mass stars amongst these 50.

\begin{figure*}
\centerline{\psfig{figure=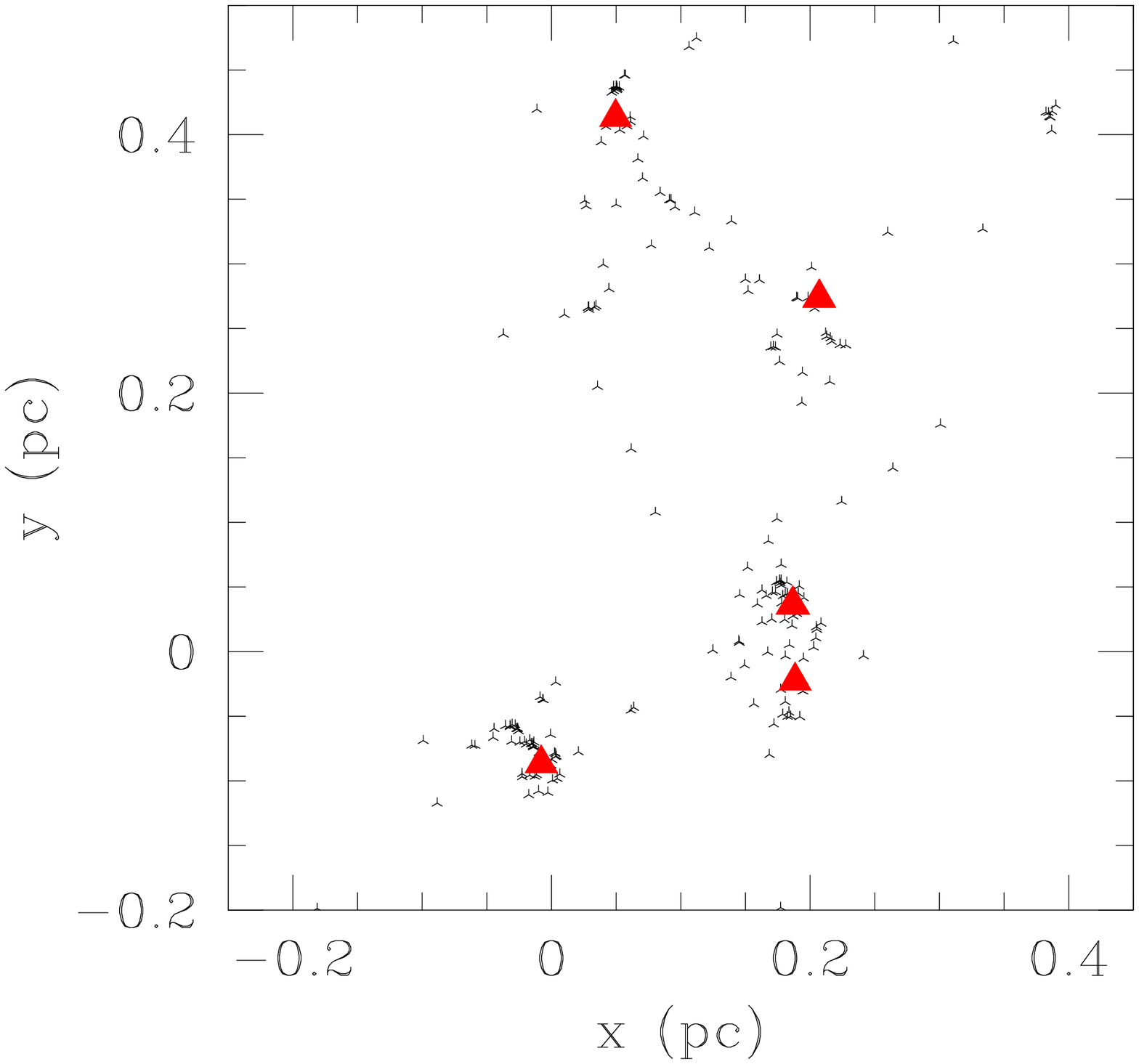,width=3.0truein,height=3.00truein}
\psfig{figure=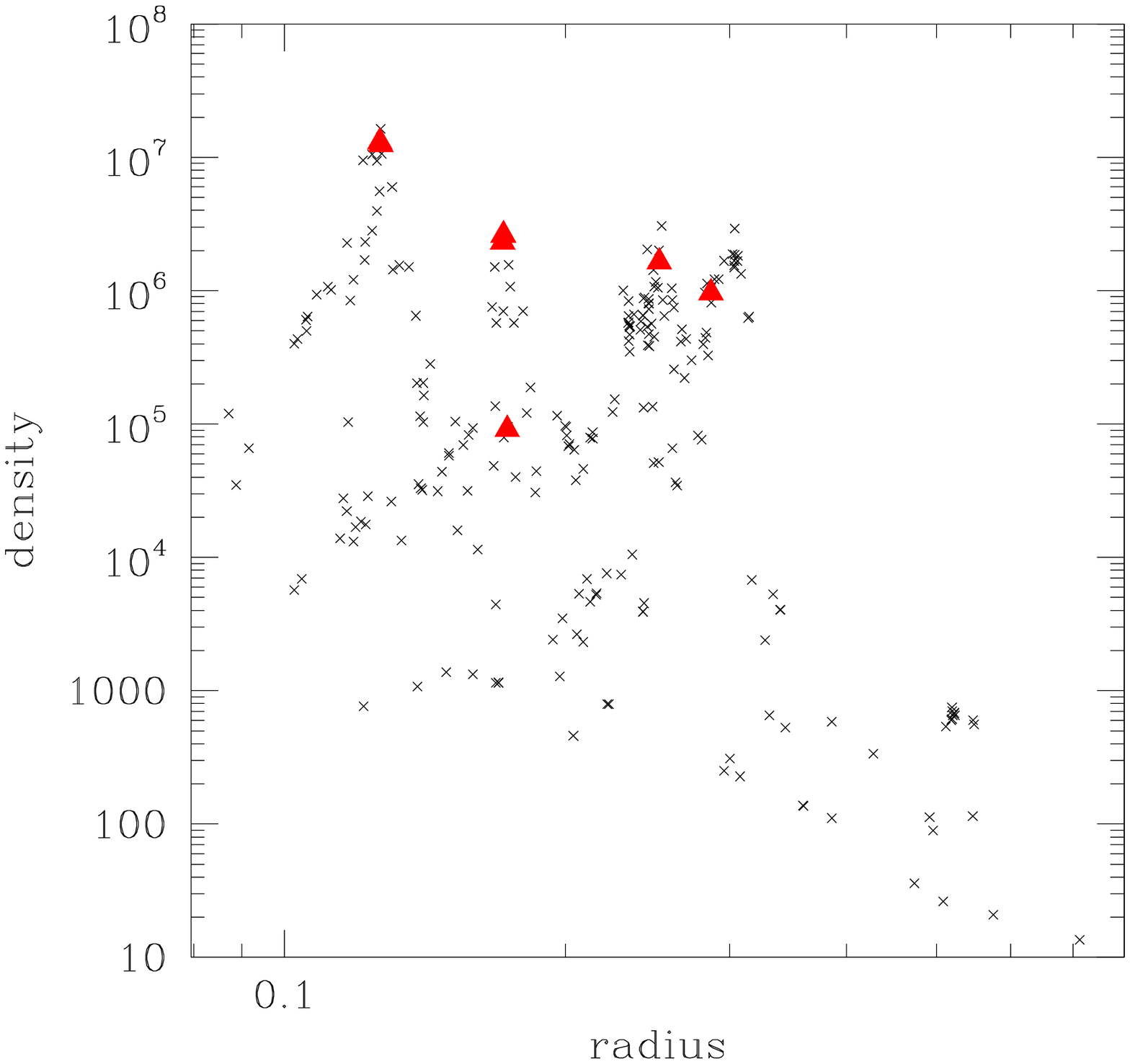,width=3.0truein,height=3.00truein}
}
\caption{\label{subclust} The positions of the stars formed by $t=1.4
\tff$ are plotted 
in the left panel, while their local stellar densities, as a function of their radius from the centre of mass, are
shown at the same time in the right panel. 
The local stellar density
is calculated as the region required to have 10 neighbours. 
Stars more
massive than 5 \solmas\ are denoted by the solid triangles. We see
from this that the more massive stars are located in the centre of
individual subclusters. }
\end{figure*}

\section{Massive stars and cluster formation}

In the previous section we have seen that massive stars accrete the
majority of their mass competitively, that is when other stars are
present in a clustered environment and the gas infalls from outside
the cluster (Figs.~\ref{massaccprof}~and~\ref{acchist}). Shock
dissipation results in a decrease in the turbulent support in the
cloud (Mac Low \etal~1998; Ostriker \etal~2001), allowing the gas to infall into the
potential wells formed by the stellar clusters. This gas accretion
into the cluster, and ultimately onto the most massive star, must also
be accompanied by any stars that form in this infalling gas
(Fig~\ref{acchist}). This then links the formation of the stellar
cluster and the formation of the massive stars therein.

As reported in Bonnell \etal~(2003), the simulation produced a number
of subclusters which evolved independently before merging to form the
final cluster. This provides multiple sites to assess how the
formation of a stellar cluster is linked to the formation of the
massive stars the clusters contain.  Figure~\ref{subclust} shows the
subclustering at $t=1.4 \tff$, roughly half-way through the simulation
and at a time where 244 stars have formed. We see that the system is
highly subclustered and that there is a relatively massive star ($m\ge
5$ \solmas) in the centre of each subcluster. In this section, we
explore the interelation between massive star formation and the
formation of a stellar cluster.

\begin{figure}
\centerline{\psfig{figure=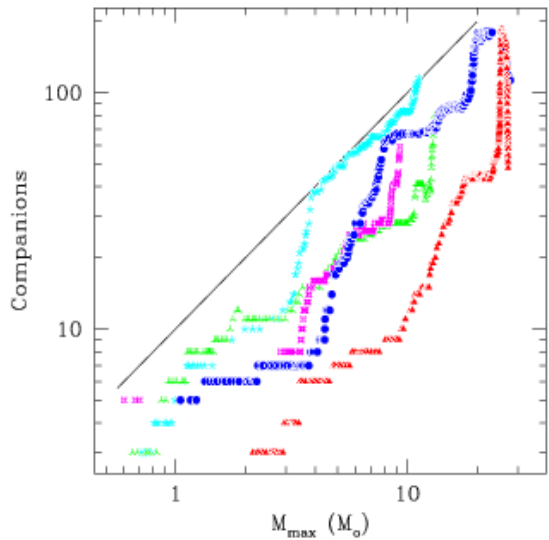,width=3.0truein,height=3.00truein}}
\caption{\label{indcvsmass} The evolution of five different stars,
that are each, at least temporarily, the most massive star in their
subcluster of size $0.1$ parsecs, is shown in the plane of the number
of companions vs the mass of the most massive star.}
\end{figure}

For what follows, we define a cluster size of $0.1$ parsecs radius.
We then evaluate, for each star, how many other stars are in this
region and whether the star considered is the most massive of all its
companions. In Figure~\ref{indcvsmass}, we plot the evolution of 
five stars which spend significant amounts of time as the most massive star
in their subcluster. The number of companion stars within $0.1$ parsecs
is plotted against the mass of this most massive star. 
Stars start out with low masses and evolve to higher masses due to
gas accretion. When new, lower-mass companions enter within $0.1$ parsecs,
the star moves upwards towards having more companions.  If a
higher-mass companion enters this volume, then the original star being
considered is no longer the most massive star in its group and is then
no longer plotted.

We see that the general evolution is from low-mass stars with few
companions towards high-mass stars with a hundred or more
companions. This tells us that as the most massive star grows by
accreting the gas that infalls onto the cluster
(Fig.~\ref{massaccprof}), the sub-cluster grows by gaining more stars.
Figure~\ref{indcvsmass} shows a nearly linear relation ($M_{max}
\propto N_{comp}$) between the number of companions and the mass of
the most massive star.  Thus, lower-mass stars are not the most
massive stars in rich clusters, nor are higher-mass stars generally
found in isolation or in sparsely populated clusters. Indeed, no
stars more massive than $\approx 5 \solm$ are found in relative
isolation ( $< 0.1$ parsecs of another star) in the simulation.
There are a number of stars that form in relative isolation, but they
never attain a significantly high mass.

The underlying explanation for the correlation evident in
Figure~\ref{indcvsmass} between the number of stars in the cluster and
the mass of the most massive star is simply that both are due to the
same process. Gas infalls onto the local potential minimum, which is
the stellar cluster. This gas forms a reservoir of mass from which the
stars accrete competitively. The most massive star 'wins' the
competition due to its mass and location in the centre of the
subcluster (Bonnell \etal~2001a). At the same time, the gas is forming
additional stars, which then infall into the cluster, increasing the
cluster numbers. Some stars actually form inside our nominal cluster
radius from the infalling gas. The infalling and newly formed stars
are generally lower-mass stars maintaining a low median stellar mass
($\approx 0.5$ \solmas).  The correlation must then arise as the star
formation is inefficient but converts an approximately constant
mass-fraction of the infalling gas into stars. The remaining mass provides
the gas reservoir from which the individual stars accrete allowing some to
attain high masses.

\subsection{Cluster properties and the IMF}

\begin{figure}
\centerline{\psfig{figure=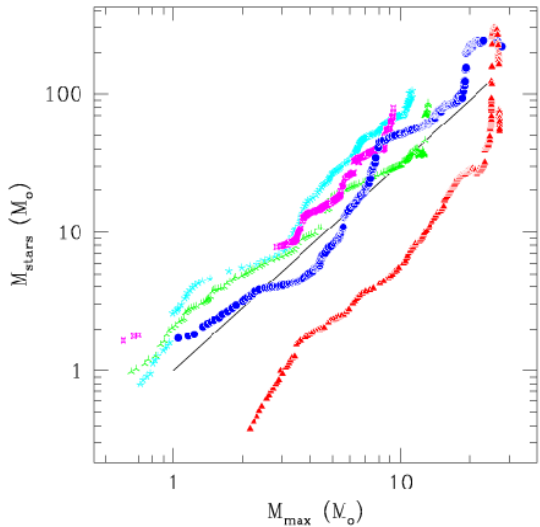,width=3.0truein,height=3.00truein}}
\caption{\label{gasacc} The total  mass in companions (within 0.1 pc)
 is plotted as a function
of the most massive star in this subcluster for the same five stars as
in Figure~\ref{indcvsmass}. Note that the companion mass does not include the
mass of the most massive star itself.  The line shows a correlation of the most
massive star's mass as a function of the mass in companion stars  as $M_*
\propto M_{\rm stars}^{2/3}$.}
\end{figure}

In this section we investigate how the mass of the most massive star
grows relative to the total mass in stars in each system. Observed
stellar clusters show a clear correlation of total mass as a function
of the mass of the most massive star in the system
(e.g. Larson~2003). This correlation could even extend out to much
higher mass systems containing intermediate and supermassive black
holes (Larson~2003; Clarke~2003). It is also of interest as one
expects that a competitive accretion model naturally results in a
similar mass spectrum independent of the size of the cluster.

For each subcluster, we evaluate the total mass in companion stars
within $0.1$ parsecs of the most massive star therein.  
We exclude the
mass of the most massive star in this total. The evolution
of the total companion mass is plotted against the most massive
star's mass in figure~\ref{gasacc}, for the five most massive stars that
spend significant amounts of time as the most massive in their systems.

Once again we see a strong correlation between these two quantities
with the general evolution of an increasing mass in companion stars as
the maximum stellar mass increases. The five systems evolve along
similar slopes in the diagram even though one of the systems is
significantly underpopulated in numbers of companion stars, and in
their total mass.  The rapid rise in companion mass in the upper
right-hand part of the diagram occurs near the end of the evolution
when the systems merge. Otherwise, the general evolution is along a
slope given by $M_{stars} \propto M_{max}^{3/2}$ or $M_{max}\propto
M_{stars}^{2/3}$. Interestingly, this result is basically what one
expects for the relation between the mass of a system and its most
massive component when they follow a Salpeter-type IMF (Larson~2003,
Clarke~2003). Thus, each individual system is evolving in a manner
consistent with populating a Salpeter-like IMF.

Each subcluster evolves by accreting stars and gas from the surrounding cloud.
The stars are typically low-mass stars of median mass $\approx 0.5$ \solmas,
similar to the clump masses from which they form. Along with the stars, gas
also infalls onto the system. 
Competitive accretion will occur such that the
more massive, slower moving, and more centrally located stars benefit
and accrete mass at higher rates (Bonnell \etal~1997,~2001a).  Competitive
accretion in the virialised cores of clusters naturally yields a
Salpeter-like IMF for higher mass stars while subvirial infall into
the cluster yields shallower IMF for lower-mass stars (Bonnell
\etal~2001b). That this process occurs approximately independent of the
size of the system explains the general evolution found in figure~\ref{gasacc}
for the individual subclusters. 
We thus find  a direct relationship between the
process which builds clusters and the mass accretion which forms the
most massive star.  

In order to explain the high-mass end of the mass spectrum, the
infalling gas must comprise a sufficient reservoir to produce a mean
stellar mass approximately twice that of the median stellar mass. This
implies that at most half of the infalling mass can be in
gravitationally bound and collapsing clumps. This then leaves at least
half of the mass to be accreted by the individual stars of the
cluster.

\section{Discussion}

In this paper we have investigated how massive star formation occurs
in the context of one numerical simulation. This is undoubtably not
the only environment in which one could image massive star formation
occurring.  The primary assumptions of the simulation are the mean
Jeans mass [the minimum mass to be gravitationally bound] in the
cloud, the total number (1000) of such Jeans masses in the cloud, and
that the cloud is globally supported by turbulence with the power
spectrum prescribed in \S 2.  The first two were chosen in order that
the mean stellar mass be similar to that found in the field and star
forming regions as well as the Galaxy as a whole. The second
assumption is a requisite in order for fragmentation to form
sufficient numbers of stars to be a stellar cluster. The exact nature
of the turbulence is a more difficult manner, although its
justification is due to the observed line-width size relationship of
molecular clouds (Larson~1981). Changing the power spectrum can affect
the fraction of gas in clumps and hence the amount left-over for
competitive accretion (Klessen, Heitsch \& Mac Low~2000). Thus,
although this could increase the maximum mass produced by the
fragmentation, it would most likely be at the cost of the lower-mass
mass spectrum. We can therefore conclude that competitive accretion
must play a role in the formation of massive stars unless they form in
isolation (away from low-mass stars).

It is worth noting here that we have not included any radiative
feedback from the massive stars (Yorke \& Sonnhalter~2002; Edgar \&
Clarke~2003), nor have we investigated the likelihood of direct
collisions in massive star formation (Bonnell, Bate \& Zinnecker~1998;
Bonnell \& Bate~2002). Radiation pressure on dust could halt infall
for stars more massive than $10 \solm$, limiting the masses of the
most massive stars. We would still get the same trend found here for
intermediate-mass stars.  Furthermore, even with the gravitational
softenning used, stellar densities reached maximum values of $\approx
10^7$ stars/pc$^{-3}$, which is near the threshold for
collisions. Thus it is conceivable that stellar mergers may play a
role in massive star formation.

\section{Conclusions}

The formation of massive stars is intricately linked to the formation
of stellar clusters.  Using a numerical simulation of the formation of
a stellar cluster from the fragmentation of a turbulent molecular
cloud, we show that massive stars {\sl do not} owe their masses to the
pre-collapse clump masses. Massive star formation is {\sl not} just a
scaled-up version of low-mass star formation. Instead, their masses
are due to subsequent infall from outside the subcluster in which the
massive star resides.  The mass of the most massive stars is therefore
primarily due to competitive accretion in a cluster environment
(Bonnell \etal~2001a). Even relatively low-mass stars (with $m
\simgreat 0.5 \solm$ derive the majority of their mass from
competitive accretion.

The infalling gas is accompanied into the subcluster by newly formed
stars. Thus, an individual cluster grows in numbers of stars as the
most massive star increases in mass. This results in a direct
correlation similar to that observed in Herbig AeBe stars (Testi
\etal~1997,~1999; Hillenbrand~1995, see Clarke \etal~2000), and
provides a physical alternative to a probabilistic sampling from an
IMF (Bonnell \& Clarke~1999).  Lastly, we find that the mass of the
most massive star in each subsystem is linked to the total mass in
stars that the system contains. This is due to the nature of competitive
accretion in producing a Salpeter-like IMF in individual systems.

\section*{Acknowledgments}
The computations reported here were performed using the U.K. Astrophysical
Fluids Facility (UKAFF).


\begin{thebibliography}{}

\bibitem[]{} Bate M. R., Bonnell I. A., Price N. M., 1995, \mnras, 277, 362.

\bibitem[]{} Bate M. R., Burkert A., 1997, \mnras, 288, 1060.



\bibitem[\protect\citeauthoryear{Behrend \&
Maeder}{2001}]{2001A&A...373..190B} Behrend R., Maeder A., 2001, A\&A, 373,
190


\bibitem[]{} Benz W., Bowers R. L., Cameron A. G. W., Press B., 1991, \apj, 348, 647.

\bibitem[]{} Bonnell I. A., Bate M. R., 2002, \mnras, 336, 659

\bibitem[]{} Bonnell I. A., Bate M. R., Clarke C. J., Pringle J. E., 1997, \mnras, 285, 201.

\bibitem[]{} Bonnell I. A., Bate M. R., Clarke C. J., Pringle J. E., 2001a, \mnras, 323, 785.

\bibitem[]{} Bonnell I. A., Bate M. R., Vine S. G., 2003, \mnras, 343, 413 

\bibitem[]{} Bonnell I. A., Bate M. R., Zinnecker H., 1998, \mnras, 298, 93.

\bibitem[]{} Bonnell I. A., Clarke C. J., 1999, \mnras, 309, 461

\bibitem[]{} Bonnell I. A., Clarke C. J., Bate M. R., Pringle J. E., 2001b, \mnras, 324, 573.

\bibitem[]{} Clarke C.J., 2003, {\sl Carnegie Observatories Astrophysical Series I: Coevolution of Black Holes and Galaxies}, ed. L. C. Ho, (Cambridge University Press), in press

\bibitem[]{} Clarke C.J., Bonnell I. A., Hillenbrand L. A., 2000, , in {Protostars and Planets IV} (eds V. Mannings, A. P. Boss and S. Russell), 151.

\bibitem[]{} Delgado-Donat\'e E., Clarke C.J., Bate M.R., 2004, MNRAS, in press

\bibitem[\protect\citeauthoryear{Edgar \&
Clarke}{2003}]{2003MNRAS.338..962E} Edgar R., Clarke C., 2003, MNRAS, 338,
962

\bibitem[\protect\citeauthoryear{Elmegreen \&
Krakowski}{2001}]{2001ApJ...562..433E} Elmegreen B.~G., Krakowski A., 2001,
ApJ, 562, 433

\bibitem[]{} Hillenbrand L. A., 1995, PhD thesis, University of Massachusetts, Amherst.

\bibitem[\protect\citeauthoryear{Klessen \&
Burkert}{2001}]{2001ApJ...549..386K} Klessen R.~S., Burkert A., 2001, ApJ,
549, 386 

\bibitem[\protect\citeauthoryear{Klessen, Heitsch, \& Mac
Low}{2000}]{2000ApJ...535..887K} Klessen R.~S., Heitsch F., Mac Low M.,
2000, ApJ, 535, 887 

\bibitem[]{} Lada C. J., Lada E. A., 2003, ARA\&A, in press

\bibitem[\protect\citeauthoryear{Larson}{1969}]{1969MNRAS.145..271L} Larson
R.~B., 1969, MNRAS, 145, 271

\bibitem[]{} Larson R. B., 1981, MNRAS, 194, 809.

\bibitem[]{} Larson R. B., 1992, MNRAS, 256, 641

\bibitem[\protect\citeauthoryear{Larson}{2003}]{} Larson R. B., 2003, {in Galactic Star formation Across the Stellar Mass Spectrum}, ASP Conference Series, ed. J. M. De Buizer, p. 65

\bibitem[\protect\citeauthoryear{Mac Low et
al.}{1998}]{1998PhRvL..80.2754M} Mac Low M., Klessen R.~S., Burkert A.,
Smith M.~D., 1998, PhRvL, 80, 2754 



\bibitem[]{} Masunaga H., Miyama S. M.; Inutsuka, S., 1998, ApJ, 495, 346


\bibitem[]{} McKee C., Tan J., 2002, Nature, 416, 59

\bibitem[\protect\citeauthoryear{McKee \& Tan}{2003}]{2003ApJ...585..850M}
McKee C.~F., Tan J.~C., 2003, ApJ, 585, 850

\bibitem[]{} Monaghan J. J., 1992, \ARAA, 30, 543.

\bibitem[\protect\citeauthoryear{Motte, Andr\'e, \&
Neri}{1998}]{1998A&A...336..150M} Motte F., Andr\'e P., Neri R., 1998, A\&A,
336, 150 

\bibitem[\protect\citeauthoryear{Murray \& Lin}{1996}]{1996ApJ...467..728M}
Murray S.~D., Lin D.~N.~C., 1996, ApJ, 467, 728

\bibitem[\protect\citeauthoryear{Osorio, Lizano, \&
D'Alessio}{1999}]{1999ApJ...525..808O} Osorio M., Lizano S., D'Alessio P.,
1999, ApJ, 525, 808 

\bibitem[]{} Ostriker E. C., Stone J. M., Gammie C. F.,  2001, \ApJ, 546, 980.

\bibitem[\protect\citeauthoryear{Padoan \&
Nordlund}{2002}]{2002ApJ...576..870P} Padoan P., Nordlund {\AA}., 2002,
ApJ, 576, 870

\bibitem[]{} Shu F. H., Adams F. C., Lizano S., 1987, \ARAA, 25, 23


\bibitem[]{} Testi L., Palla F., Prusti T., Natta A., Maltagliati S., 1997, \AaA, 320, 159.

\bibitem[]{} Testi L., Sarget A. I., Olmi L., Onello J. S., 2000, \apjl, 540, L53 

\bibitem[\protect\citeauthoryear{Wolfire \&
Cassinelli}{1987}]{1987ApJ...319..850W} Wolfire M.~G., Cassinelli J.~P.,
1987, ApJ, 319, 850

\bibitem[\protect\citeauthoryear{Yorke \&
Sonnhalter}{2002}]{2002ApJ...569..846Y} Yorke H.~W., Sonnhalter C., 2002,
ApJ, 569, 846

\bibitem[\protect\citeauthoryear{Yorke \&
Kruegel}{1977}]{1977A&A....54..183Y} Yorke H.~W., Kruegel E., 1977, A\&A,
54, 183 

\bibitem[]{} Zinnecker H., 1982, in {\sl Symposium on the Orion Nebula to
Honour Henry Draper}, eds A. E. Glassgold \etal, New York Academy of
Sciences, p. 226

\end{thebibliography}
\end{document}